\documentclass[twocolumn,prl,10pt,noshowpacs,aps,footinbib]{revtex4-1}
\usepackage{color,times,url,graphicx,afterpage}
\usepackage{multirow,bm,braket,soul,float}
\usepackage{amsmath,xcolor,hyperref,mathtools}
\usepackage{braket}
\usepackage[normalem]{ulem}

\renewcommand{\figurename}{Figure}
\renewcommand{\tablename}{Table}
\renewcommand{\thetable}{\arabic{table}}

\begin{document}

\title{Phonon Screening of Excitons in Semiconductors: Halide Perovskites and Beyond}

\author{Marina R. Filip$^{1,2,3\dagger}$}
\author{Jonah B. Haber$^{2,\dagger}$}
\author{Jeffrey B. Neaton$^{2,4,5}$}
\email{jbneaton@lbl.gov}
\affiliation{$^1$Department of Physics, University of Oxford}
\affiliation{$^2$Department of Physics, University of California Berkeley}
\affiliation{$^3$Molecular Foundry, Lawrence Berkeley National Lab}
\affiliation{$^4$Materials Science Division, Lawrence Berkeley National Lab}
\affiliation{$^5$Kavli Energy NanoSciences Institute at Berkeley\\
$\dagger$ \rm Authors contributed equally to this manuscript.}

\begin{abstract}
         The {\it ab initio} Bethe-Salpeter equation (BSE) approach, an established method for the study of excitons in materials, is typically solved in a limit where only static screening from electrons is captured. Here, we generalize this framework to also include dynamical screening from phonons at lowest order in the electron-phonon interaction. We apply this generalized BSE approach to a series of inorganic lead halide perovskites, CsPbX$_3$, with X = Cl, Br, and I. We find that inclusion of screening from phonons significantly reduces the computed exciton binding energies of these systems. 
         By deriving a simple expression for phonon screening effects, we reveal general trends for the importance of phonon screening effects in semiconductors and insulators, based on a hydrogenic exciton model. We demonstrate that the magnitude of the phonon screening correction in isotropic materials can be reliably predicted using four material specific parameters: the reduced effective mass, the static and optical dielectric constants, and the phonon frequency of the most strongly coupled  LO phonon mode. This framework helps to elucidate the importance of phonon screening and its relation to excitonic properties in a broad class of semiconductors.
\end{abstract}
\maketitle
Excitons are central to a wide range of optoelectronic applications, from photovoltaics and photocatalysis, to light emission and lasing~\cite{Takanabe2017,  Herz2018, Ai2018, Ginsberg2020}; they emerge from the many-body interactions between charge carriers, photons, and phonons in optoelectronic materials~\cite{Knox}. In many bulk semiconductors, weakly bound Wannier-Mott excitons can be understood with a hydrogenic model~\cite{Wannier, Elliot}, in which the attractive Coulomb interaction between a photoexcited electron-hole pair is screened by a dielectric constant ${\varepsilon}$. In this picture, the exciton binding energy is $\mu/(2{\varepsilon^2})$ in atomic units, where $\mu$ is the magnitude of the reduced effective mass of the electron-hole pair~\cite{Wannier}. Optical measurements under high magnetic fields use this model to extract the exciton binding energy, $E_B$ and $\mu$~\cite{Miyata2015, Makado1986}. In ionic or multicomponent semiconductors, an ``effective dielectric constant'', $\varepsilon_{\rm eff} = \sqrt{2E_B/\mu}$, is frequently reported, usually taking values between the optical, $\varepsilon_\infty$, and static, $\varepsilon_0$, dielectric constants. The use of $\varepsilon_{\rm eff}$ approximately accounts for the fact that the electron-hole interaction is screened by both the electrons and phonons~\cite{Mahanti1970, Mahanti1972, Herz2018}. However, it also obscures the details of specific phonons contributing to $\varepsilon_{\rm eff}$, and it does not explain whether or why electron or phonon screening might be important in a given case. Rigorous {\it ab initio} calculations would therefore be of great value in this context.

{\it Ab initio} many-body perturbation theory calculations within the $GW$ approximation~\cite{Hedin, Hybertsen} and the Bethe-Salpeter equation (BSE)~\cite{Rohlfing1998, Albrecht1998} approach have been successful in quantitatively understanding the quasiparticle band structure and optical excitations of materials ranging from the simplest III-V semiconductors~\cite{Rohlfing2000} to materials with heavy elements \cite{Malone2013} or hybrid organic-inorganic components~\cite{Filip2014-2}, low dimensionality~\cite{Qiu2013}, and intrinsic defects~\cite{Refaely-Abramson2018}. First principles methods including the effects of lattice vibrations have led to new understanding of the renormalization of the electronic band structure due to electron-phonon interactions~\cite{Giustino2017,Giustino2010,Antonius2014}, as well as optical absorption~\cite{Kioupakis,Zacharias} and photoluminescence lineshapes~\cite{Marini, Antonius}.%

Recently, several first principles studies of a broad range of materials predicted exciton binding energies which are overestimated with respect to experiment~\cite{Bechstedt2005, Fuchs2008, Schleife2018, Bechstedt2019, Bokdam2016, Umari2018}. In particular,
Ref.~\cite{Bokdam2016} recently reported calculated exciton binding energies of hybrid organic-inorganic lead-halide perovskites  which overestimate experimental measurements by up to a factor of 3. Ref.~\cite{Bokdam2016} attributed this overestimation to the coupling of the constituent free electrons and holes to phonons (hereafter referred to as `polaronic effects').
On the other hand, Ref.~\cite{Umari2018} used an approximate model dielectric function to conclude that phonon screening due to infrared active phonons renormalizes the exciton binding energy by up to 50\%, bringing calculated values in much closer agreement with experiment. Since both reports are based on approximate hypotheses and implementations of phonon effects, it is not yet clear how these conclusions may be reconciled, in the absence of a complete {\it ab initio} calculation. 
The problem of electrons and holes interacting in a phonon field has been studied using phenomenological models, assuming parabolic electronic band structure and a phonon spectrum consisting of a single dispersionless phonon~\cite{Mahanti1970,Mahanti1972, Pollmann1977, Kane1978, Matsuura1980}. 
However, rigorous inclusion of polaronic and phonon screening effects within the BSE formalism remains an open challenge. 

In this Letter, we extend the standard {\it ab initio} BSE formalism to include phonon screening effects at lowest order in the electron-phonon interaction. We introduce an additive, {\bf q}- and $\omega$-dependent contribution to the screened Coulomb interaction, $W$, associated with phonons, adopting a general form developed by Hedin and Lundquist~\cite{Hedin1970} but neglected in contemporary calculations. We apply this framework to a set of all-inorganic lead-halide perovskite crystals in the low temperature, orthorhombic phase using the {\it ab initio}  Fr\"{o}hlich electron-phonon vertex introduced in Ref.~\cite{Verdi2015}, and we show that phonon screening plays a major, but not exclusive, role in the exciton binding energies of this emergent class of optoelectronic materials. Finally, we develop a simple but general expression for the phonon-screened exciton binding energy for arbitrary isotropic semiconductors in terms of $\mu$, $\varepsilon_{\infty}$, $\varepsilon_0$, and $\omega_{\rm LO}$, providing a means for identifying semiconductors for which phonon screening effects will be significant.

\begin{figure*}[t!]
        \begin{center}
                \includegraphics[width=0.85\textwidth]{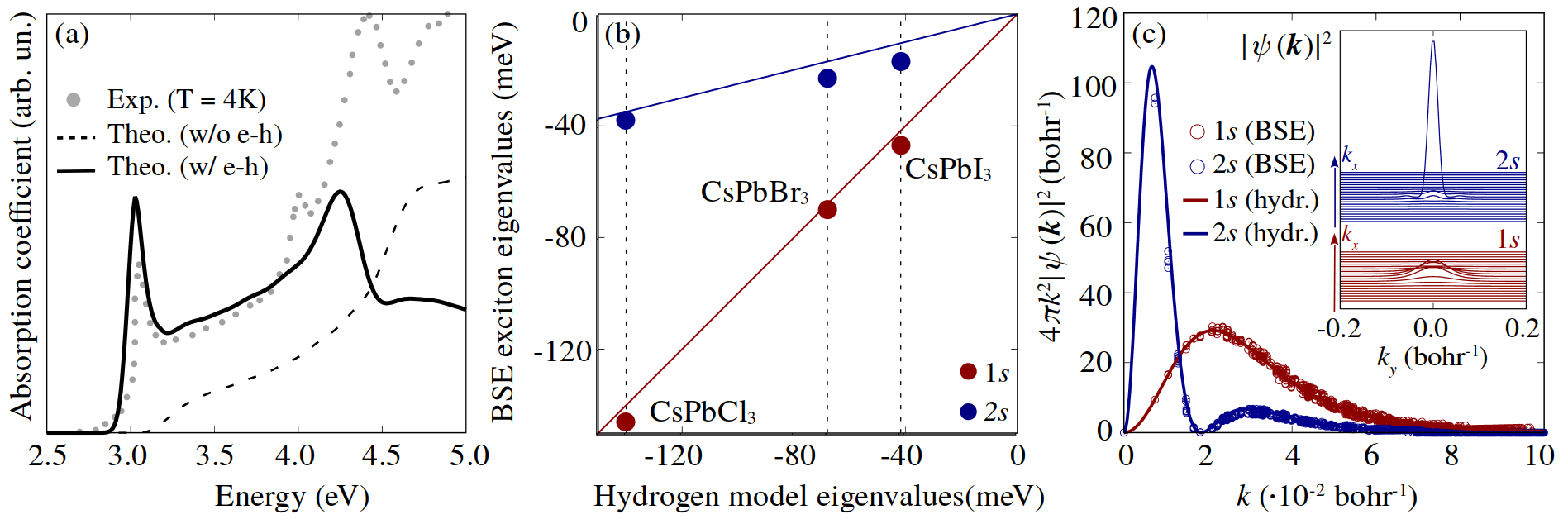}
                \vspace{-0.40cm}
                \caption{(a) Optical absorption spectrum calculated within $GW$/BSE (continuous line), RPA (dotted line), and from experiment (grey dots)~\cite{Heindrich1978} for CsPbCl$_3$. Calculated spectra are blue-shifted by 0.3~eV to match the experimental onset from Ref.~\citenum{Heindrich1978}. See SI for similar spectra for CsPbBr$_3$ and CsPbI$_3$ ~\cite{si}
                (b) Exciton binding energies predicted from $GW$/BSE (filled circles) and the hydrogen model (lines).
                (c) Exciton radial probability density (main) and  probability of localization (inset) in reciprocal space, calculated from $GW$/BSE and the hydrogen model for 1s (dark red) and 2s (dark blue) states.
                \label{fig1}\vspace{-0.5cm}}
        \end{center}
        \vspace{-0.5cm}
\end{figure*}
    
In the standard {\it ab initio} reciprocal-space $GW$-BSE approach, the BSE can be written, in the Tamm-Dancoff approximation~\cite{Rohlfing1998, Rohlfing2000}, as
{\small\ \begin{eqnarray}\vspace{-0.5cm}
\label{eq0}
        \Delta_{c\mathbf{k}v\mathbf{k}}A_{cv\mathbf{k}}^S + \sum_{c'v'\mathbf{k'}}
        K_{cv\mathbf{k},c'v'\mathbf{k'}}(\Omega_S) A_{c'v'\mathbf{k'}}^S  
        =\Omega_S A_{cv\mathbf{k}}^S,
        \vspace{-0.5cm}
\end{eqnarray}}\noindent
where $\Delta_{c\mathbf{k}v\mathbf{k}} = E_{c\mathbf{k}} - E_{v\mathbf{k}}$, with $E_{c\mathbf{k}}$ and $E_{v\mathbf{k}}$ the quasiparticle energies of the free electron and hole with band indices and wavevectors $c\mathbf{k}$ and $v\mathbf{k}$, respectively, usually calculated within the $GW$ approximation~\cite{Hedin, Hybertsen}. Exciton energies and expansion coefficients, in the electron-hole basis, are given by $\Omega_S$, and $A_{cv\mathbf{k}}^S = \braket{cv\mathbf{k}|S}$ respectively, with $S$ the principal quantum number for the exciton, and $\ket{cv\mathbf{k}}$  the product state of an electron-hole pair, where the components of the products are typically Kohn-Sham wave functions computed with density functional theory (DFT)~\cite{Hohenberg}. 

The electron-hole kernel, $K$, couples products of the single-particle states and is, at lowest order, written as the sum of two terms, a repulsive exchange term, $K^{\rm x}$, which is negligible for weakly bound excitons~\cite{Strinati}, and an attractive direct term, $K^{\rm D}$, given by, as in Ref.~\cite{Strinati},
{\vspace{-0.20cm}\small\ \begin{align}\label{eq1}
        K^{\rm D}_{cv\mathbf{k},c'v'\mathbf{k'}}(\Omega) = -\Bigg \langle cv \mathbf{k}\Bigg |
        \frac{i}{2\pi}\int d\omega~e^{-i\omega \eta} W(\mathbf{r}, \mathbf{r'};\omega)\times&  \\\nonumber
                 \Bigg[\frac{1}{\Omega-\omega-\Delta_{c'\mathbf{k'}v\mathbf{k}}+i\eta} +
                \frac{1}{\Omega+\omega-\Delta_{c\mathbf{k}v'\mathbf{k'}}+i\eta}\Bigg] & \Bigg|
                c'v' \mathbf{k'}\Bigg \rangle, \nonumber
                \vspace{-0.20cm}
\end{align}}\noindent
where $\eta$ is a positive infinitesimal quantity, and $W(\mathbf{r}, \mathbf{r'}; \omega)$ is the time-ordered screened Coulomb interaction, which typically only includes electronic contributions to screening. In general, the BSE must be solved self-consistently, as $K^D$ depends on $\Omega_S$.

As discussed by Hedin and Lundquist~\cite{Hedin1970}, $W$ can rigorously be written as the sum of an electronic, $W^{\rm el}$, and ionic (or phonon), $W^{\rm ph}$ part, i.e., $W(\mathbf{r}, \mathbf{r'}; \omega) = W^{\rm el}(\mathbf{r}, \mathbf{r'}; \omega)+ W^{\rm ph}(\mathbf{r}, \mathbf{r'}; \omega)$. In standard BSE calculations, $W^{\rm ph}$ is ignored while $W^{\rm el}$ is routinely computed within the random-phase approximation (RPA)~\cite{Adler, Wiser}, neglecting the frequency dependence.
The $W^{\rm ph}$ term may be written in the form (see SI~\cite{si}) 
{\small \vspace{-0.1cm}\begin{equation}\label{wph}
W^{\rm ph}(\mathbf{r},\mathbf{r'};\omega) = \sum_{\mathbf{q}\nu}D_{\mathbf{q}\nu}(\omega)g_{\mathbf{q}\nu}(\mathbf{r})g^*_{\mathbf{q}\nu}(\mathbf{r'}),
\vspace{-0.20cm}
\end{equation}}\noindent
where $D_{\mathbf{q}\nu}(\omega)$ is the phonon propagator and $g_{\mathbf{q}\nu}(\mathbf{r})$ is the electron-phonon vertex, encoding the probability amplitude for an electron at $\mathbf{r}$ to scatter off a phonon with crystal momentum $\mathbf{q}$ and branch index $\nu$ (see SI~\cite{si})~\cite{Hedin1970}.

Incorporating $W^{\rm ph}$ into  the BSE kernel, we obtain the phonon contribution to the real part of the direct electron-hole kernel matrix elements as follows (written here in the exciton basis; see SI for details~\cite{si}):
{\small \vspace{-0.1cm}\begin{eqnarray}\label{kph}
            \hspace{-4mm}{ {\rm Re}[K^{\rm ph}_{SS'}(\Omega)] = - \hspace{-1mm}{\sum_{\substack{cv\mathbf{k} \\ c'v'\mathbf{k'} \nu}} 
            A^{S*}_{cv\mathbf{k}}g_{cc'\nu}(\mathbf{k'},\mathbf{q})g_{vv'\nu}^*(\mathbf{k'},\mathbf{q})A_{c'v'\mathbf{k'}}^{S'}}} \\ \times  \bigg[\frac{1}{\Omega-\Delta_{c'\mathbf{k'}v\mathbf{k}}-\omega_{\mathbf{q}\nu}}+\frac{1}{\Omega-\Delta_{c\mathbf{k}v'\mathbf{k'}}-\omega_{\mathbf{q}\nu}}\bigg], \nonumber
        \end{eqnarray}}\noindent
where $g_{nm\nu}(\mathbf{k'},\mathbf{q}) = \braket{m\mathbf{k}'+\mathbf{q}|g_{\mathbf{q}\nu}|n\mathbf{k}'}$, with $\mathbf{q}=\mathbf{k}-\mathbf{k}'$. 
From Brillouin-Wigner perturbation theory, it follows that the change in the exciton energy, $\Delta \Omega_S$, due to phonon screening, is related to $K^{\rm ph}$ through  $\Delta \Omega_S  = {\rm Re}[K_{SS}^{\rm ph}(\Omega_S + \Delta \Omega_S)]$, in the limit where off-diagonal components of $K^{\rm ph}$ can be neglected. 

We pause to note that $W^{\rm ph}$ should, in principle, be included in both the BSE kernel and $GW$ self-energy. The contribution to the latter, i.e., $iGW^{\rm ph}$, is equivalent to the Fan-Migdal electron-phonon self-energy~\cite{Giustino2017}, and leads to polaronic mass enhancement and energy renormalization (e.g.~\cite{Schlipf2019}) effects that would naively tend to increase the exciton binding energy over the bare or phonon-screened values. However as discussed in Ref.~\cite{Mahanti1970}, interference between electron and hole polaron clouds upon overlap (hereafter referred to as ``interference effects'') can counter mass enhancement effects, reducing the overall binding energy. A full {\it ab initio} study of bound electron-hole polarons, including the competition between mass enhancement and interference effects, as described by higher-order or self-consistent terms in the BSE kernel, requires a separate study and is beyond the scope of this work; thus, we restrict our focus here to quantifying and understanding the phonon screening contribution to the exciton binding energy, building on prior work~\cite{Bechstedt2019, Umari2018} and the standard $GW$ approximation in all cases.

We now apply Eqs.~\ref{eq0} and~\ref{eq1}, as implemented in the BerkeleyGW code~\cite{BGW}, to CsPbX$_3$ lead halide perovskites, with X = Cl, Br, I. In Table~\ref{tb1} we compare calculated $G_0W_0$ band gaps and reduced effective masses to experiment. The computed gaps consistently underestimate experiment by up to 0.5 eV (see Table S2 of the SI~\cite{si}), a shortcoming of one-shot $G_0W_0$ approximation previously identified in a number of computational studies~\cite{Brivio2014, Filip2014-2,Leppert2019,Wiktor2017}. Furthermore, the reduced effective masses of CsPbI$_3$ and CsPbBr$_3$ agree well with recent magneto-optical measurements at high magnetic fields, while for CsPbCl$_3$ the reduced mass is slightly underestimated with respect to experiment~\cite{Miyata2015, Baranowski2020}. In the same table we also report exciton binding energies calculated within the standard BSE approach, including only electronic screening when constructing the electron-hole kernel. In agreement with previous calculations~\cite{Bokdam2016,Umari2018}, we find that exciton binding energies neglecting phonon screening overestimate experiment by up to a factor of 3. Despite these discrepancies, after blue-shifting the calculated optical absorption spectrum to align with experiment, we find the lineshape to be in good agreement with measurements at low temperature (Figure~\ref{fig1}a for CsPbCl$_3$ and Figure S2 for CsPbBr$_3$ and CsPbI$_3$). 

We further observe that low-lying optical excitations are well described using a Mott-Wannier hydrogen model.
In Figure~\ref{fig1}b we compare the BSE solutions for the $1s$ and $2s$ excitonic states with those predicted by the hydrogen model with $\mu$ calculated from $G_0W_0$ band structure, and with $\varepsilon_\infty$  calculated within the RPA~\cite{Adler,Wiser}. We find a maximum difference between the hydrogenic model and the standard BSE calculations of 6~meV for both $1s$ and $2s$ excitonic energies across all three halide perovskites. Furthermore, in Figure~\ref{fig1}c, we find that the excitonic wave functions calculated with BSE are accurately described by the hydrogenic model. 

\begin{table*}[t!]
        \begin{center}
        \begin{tabular}{ l c c c c c c c c c c c c c c c c c c c c c c c c c c}
                        \hline \hline
                                    & & $\omega_{\rm LO}$ (meV) & & $\omega_{\rm LO}^{\rm exp}$ (meV)
                                    & & $E_{\rm B}$ (meV)  & &  $\Delta E_B$ (meV)   &&
                                    $E^{\rm exp}_{\rm B}$ (meV) & & $\mu$ ($m_{\rm e}$) &&
                                    $\mu^{\rm exp}$ ($m_{\rm e}$) & & $\varepsilon_{\infty}$ & &
                                    $\varepsilon_{\infty}^{\rm exp}$
                                    && $\varepsilon_0$ && $\varepsilon_0^{\rm exp}$ \\
                        \hline
                        CsPbCl$_3$ && 26 && 25.3/28.0$^{\rm a}$; 27.5$^{\rm b}$ &&  146 && -17  &&
                                      72$\pm$ 3$^{\rm c}$; 64$\pm$1.5$^{\rm d}$  && 0.142 && 0.202$\pm$0.01$^{\rm d}$ & & 3.7 & & 3.7$^{\rm a}$ & &
                                      17.5 & & 15.7$^{\rm a}$ \\
                        CsPbBr$_3$ && 18 && 17.9/20.4$^{\rm e}$ && 70  && -12   && 
                        33$\pm$ 1$^{\rm f}$;38 $\pm$ 3$^{\rm c}$  && 0.102 && 0.126$\pm$0.02$^{\rm f}$  & &
                                      4.5 & & N/A && 18.6 && N/A\\
                        CsPbI$_3$  && 14 && 14.2$^{\rm g}$      && 47  &&  -8   && 15 $\pm$ 1$^{\rm f}$ &&
                        0.093             && 0.114$\pm$0.01$^{\rm f}$  & & 5.5 & & N/A && 22.5 && N/A \\
                        \hline \hline
                \end{tabular}
                 \vspace{-0.4cm}
                \caption{Calculated LO phonon frequencies ($\omega_{\rm LO}$),
                bare exciton binding energies ($E_{\rm B}$), phonon screening corrections ($\Delta E_B$), reduced effective masses ($\mu$), static ($\varepsilon_0$ from DFPT) and optical dielectric constants ($\varepsilon_{\infty}$ from DFPT and $G_0W_0$),
                and corresponding experimental data from Refs.
                $^{\rm a}$\cite{Nedelec2003}; 
                $^{\rm b}$\cite{Wakamura2001}; 
                $^{\rm c}$\cite{Zhang2016};
                $^{\rm d}$\cite{Baranowski2020-1};
                $^{\rm e}$\cite{Iaru2017};    
                $^{\rm f}$\cite{Yang2017};     
                $^{\rm g}$\cite{Zhao2019}.
                \label{tb1}}
      \end{center}
        \vspace{-0.9cm}
\end{table*}

We now investigate how including phonon screening contributions shifts the energy of the lowest bound exciton by explicitly computing $K^{\rm ph}$. We make two approximations to Eq~\ref{kph}: we use the analytic hydrogenic expressions for the exciton coefficients $A_{cv\mathbf{k}}^S$, and we approximate the electron-phonon matrix elements using a multi-mode, {\it ab initio} Fr\"ohlich vertex, introduced in Ref.~\cite{Verdi2015}:
{\small\vspace{-0.2cm}\begin{equation}\label{verdi}
g_{\mathbf{q} \nu}=i\frac{4\pi}{V}\sum_{\kappa}\Bigg(\frac{1}{2NM_\kappa\omega_{\mathbf{q}\nu}}\Bigg)^{1/2}\frac{\mathbf{q}\cdot\mathbf{Z_{\kappa}}\cdot\mathbf{e}_{\kappa\nu}(\mathbf{q})}{\mathbf{q}\cdot\mathbf{\varepsilon_\infty}\cdot\mathbf{q}},
\vspace{-0.1cm}
\end{equation}}
where $V$ is the unit cell volume, $M_\kappa$ are the atomic masses, $\mathbf{Z_\kappa}$ Born effective charge tensor and $\mathbf{e}_{\kappa\nu}(\mathbf{q})$ are the eigenvectors corresponding to the phonon modes $\omega_{\mathbf{q}\nu}$ for each atom indexed by $\kappa$. 
With the above simplifications, Eq.~\ref{kph} becomes:
{\small\begin{align}\label{kph1} \vspace{-0.3cm}
&\Delta \Omega_{S} = -\frac{8 a_0^3}{\pi^2} \sum_{\mathbf{kq}\nu} \frac{|g_{\mathbf{q}\nu}|^2}{[1+a_0|\mathbf{k}|^2]^2[1+a_0^2|\mathbf{k}+\mathbf{q}|^2]^2} \times \\ \nonumber
&  \bigg[\frac{1}{\Omega_{S} - \Delta_{c\mathbf{k}v'\mathbf{k+q}}-\omega_{\mathbf{q}\nu}} + \frac{1}{\Omega_{S}-\Delta_{c'\mathbf{k+q}v\mathbf{k}}-\omega_{\mathbf{q}\nu}} \bigg],\vspace{-0.2cm}
\end{align}}
where $a_0$ is the exciton Bohr radius. In principle, $\Omega_{S}$ appearing in the energy denominator above should be replaced with $\Omega_{S}+\Delta \Omega_S$ and the equation should be solved self consistently. In practice, for  CsPbX$_3$ we find the above expression differs  by less than 1~meV from the  self-consistent solution justifying a ``one-shot'' approach. Finally, by definition, the change in the exciton binding energy is $\Delta E_B =-\Delta \Omega_S$.  

 The standard BSE exciton binding energies and phonon screening corrections are summarized in Table~\ref{tb1}  for all three CsPbX$_3$ perovskites. We find that phonon screening contributes to the reduction of the exciton binding energy between 12\% and 17\% for the CsPbX$_3$ series, improving the agreement with measurements reported in Refs.~\cite{Yang2017, Zhang2016, Baranowski2020-1}. However, for CsPbI$_3$, our calculated  relative phonon screening correction of 17\% is less than half of the 50\% correction predicted in Ref.~\cite{Umari2018}; as we show in the following, this discrepancy can be attributed to electronic band dispersion contributions, accounted for here but neglected in prior work.
 
To further investigate the contribution of phonon screening to the exciton binding energy, we perform a spectral decomposition on the phonon kernel (see Figure S4 of the SI~\cite{si}). For all three halide perovskites (see SI~\cite{si}), we find that the contribution of the highest lying IR active phonons accounts for more than 90\% of the expectation value of  $K^{\rm ph}$, with the remaining contribution due to the lower energy LO modes. Furthermore, as shown in Figure~S4, the phonon kernel drops sharply outside of the $\mathbf{q}\xrightarrow[]{}0$ range, a trend attributed to the strong localization of the exciton wave function around the center of the Brillouin zone, and the fast decay of the long-range electron-phonon vertex in reciprocal space.

Given the flat profile of the optical phonon band shown in Figure. S4, we can further simplify the phonon kernel by replacing the phonon frequencies with $\omega_{LO}$, and approximating the electron-phonon vertex in Eq.~\ref{kph1} using the Fr\"ohlich model~\cite{Frohlich},
   $|\displaystyle{ g^{\rm F}_{\mathbf{q}}|^2 = 4 \pi\omega_{\rm LO}/(2N V) (\varepsilon^{-1}_{\infty} - \varepsilon^{-1}_0) / q^2}$,
where $N$ is the total number of unit cells in the crystal. This approximation yields a change in the phonon screening correction of $\sim$1\% with respect to the {\it ab initio} result, indicating that the single dispersionless phonon model is a suitable approximation for  the phonon kernel in these systems. Assuming isotropic and parabolic electronic band dispersion, Eq.~\ref{kph1} can be solved analytically (see the SI for details~\cite{si}), obtaining:
{\vspace{-0.10cm}\small\ \begin{equation}
        \label{eq:screen}
        \Delta E_B = -2\omega_{LO} \Bigg(1-\frac{\epsilon_{\infty}}{\varepsilon_0}\Bigg) \frac{\sqrt{1+\omega_{LO}/E_{B}}+3}{\big(1+\sqrt{1+\omega_{LO}/E_{B}}\big)^3}.
        \vspace{-0.10cm}
\end{equation}}\noindent
For isotropic semiconductors, Eq.~\ref{eq:screen} yields very close agreement with the numerical result (see Table S4 of the SI~\cite{si}).

Since the exciton wave function is highly localized at the center of the Brillouin zone (see Fig.~S4 of the SI~\cite{si}), it is tempting to assume that the dispersion of the electronic band structure may also be neglected. This approximation leads to an even simpler expression for the change in the exciton binding energy, $\displaystyle{\Delta E_B = -2E_B \frac{\omega_{LO}}{\omega_{LO}+E_B}(1-\epsilon_\infty/\epsilon_0)}$ (see SI~\cite{si}); however, we find that it overestimates the magnitude of the phonon screening contribution by up to 50\% with respect to the {\it ab initio} result for these systems. 

To examine phonon screening trends across a wide range of semiconductors and insulators, we plot the phonon kernel relative to the bare exciton binding energy $E_B$, $|\Delta E_B|/E_{\rm B}$, as a function of $E_{\rm B}/\omega_{\rm LO}$, and $\varepsilon_{0}/\varepsilon_{\infty}$, in Figure~\ref{fig3}, following Eq.~\ref{eq:screen}. We overlay our calculations for the CsPbX$_3$ series, as well as some other isotropic semiconductors and insulators such as CdS, GaN, AlN and MgO (see SI for computational details~\cite{si}). In all cases considered, the inclusion of phonon screening effects reduces the exciton binding energy significantly, bringing calculated values in closer agreement with experiment.

Particularly for halide perovskites, our calculations reconcile prior reports, and clearly establish the importance of phonon screening effects for excitons in halide perovskites, in agreement with Ref.~\cite{Umari2018}. However, corrections due to phonon screening do not fully account for the discrepancy between calculated and measured exciton binding energies. Considering the systematic overestimation of exciton binding energies for all systems beyond halide perovskites, we expect that the net contribution of polaronic mass enhancement~\cite{Schlipf2019} and interference effects~\cite{Mahanti1970} will further reduce the exciton binding energies  and  improve the agreement with experiment, as proposed by Ref.~\cite{Mahanti1970, Mahanti1972} for MgO and several other semiconductors. However, $G_0W_0$-BSE calculations of halide perovskites are known to exhibit a strong dependence to the mean-field starting point~\cite{Leppert2019}, and the electron-phonon matrix elements, computed  starting from the standard Kohn-Sham eigensystem may underestimate couplings obtained from higher level theory~\cite{Laflamme2010,Antonios2011,Li2019}. Therefore, a detailed benchmarking of these effects  is required, in addition to simply including polaronic effects. While we reserve this detailed analysis to future studies, we emphasize that the relative phonon screening correction derived in this study is robust, and the formalism introduced here is independent of the choice of computational setup. 

\begin{figure}[b!]
\vspace{-0.5cm}
      \begin{center}              
            \includegraphics[width=0.42\textwidth]{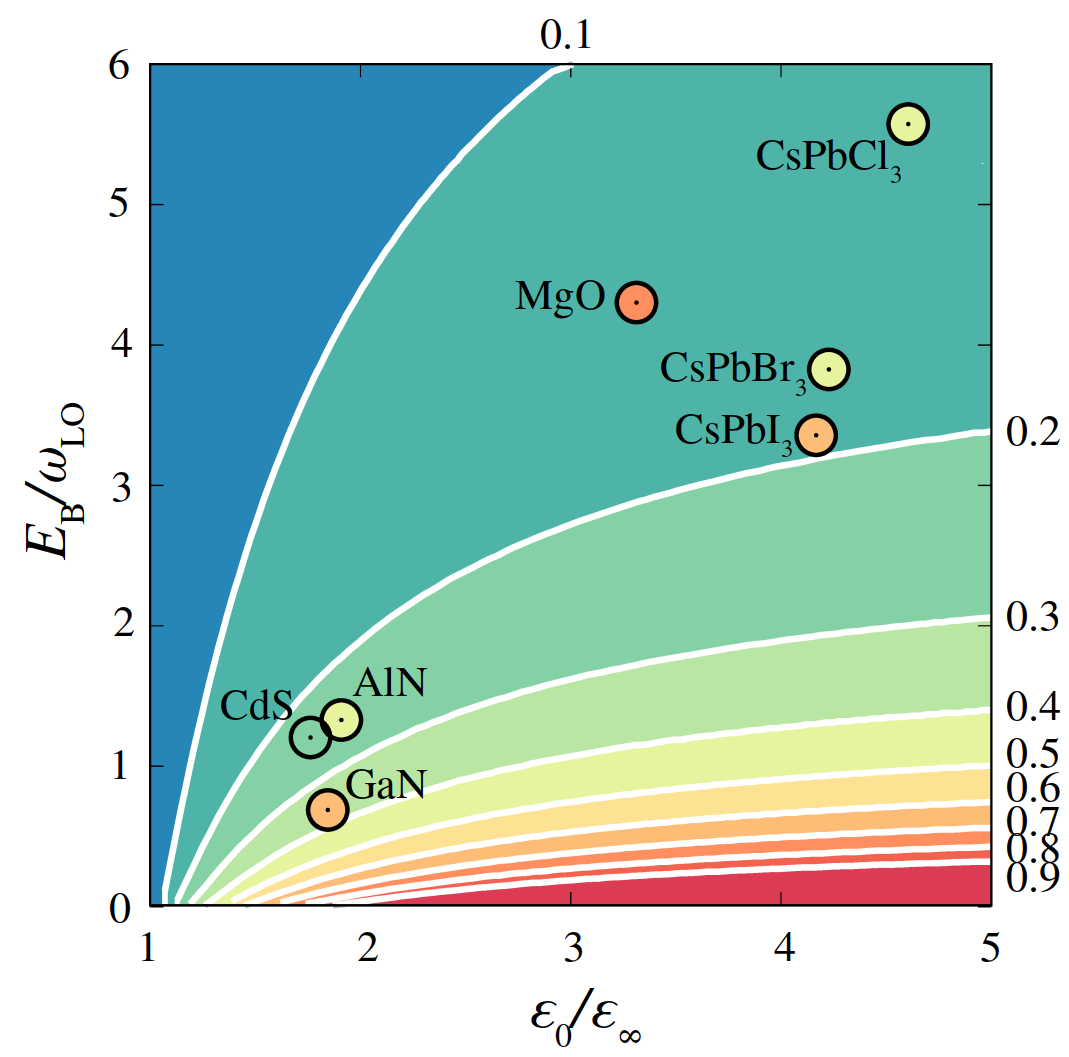}
              \vspace{-0.4cm}
              \caption{
              Color map of $\Delta E_{\rm B}/E_{\rm B}$, calculated using  Eq.~\ref{eq:screen},
              as a function of $\varepsilon_0/\varepsilon_\infty$ and $E_{\rm B}/\omega_{\rm LO}$.
              The isoline values are marked at the upper and rightmost edge of the plot. 
               The color of each circle corresponds to the ratio $(E_B-E_B^{\rm exp})/E_B$, as read on the color map. Calculated and experimental exciton binding energies are summarized in Table S4.
       \label{fig3}}
       \end{center}
       \vspace{-0.5cm}
\end{figure}

As a general trend, Figure~\ref{fig3} highlights that the magnitude of the phonon screening correction increases as the ratio $E_B/\omega_{LO}$ decreases, and in systems with a large static dielectric constant. Further, all parameters appearing in Eq.~\ref{eq:screen} and depicted in Figure~\ref{fig3}  can be readily computed or measured experimentally so that this simplified picture can be used in both theoretical and experimental contexts to directly assess the expected phonon screening correction to the bare exciton binding energy, and identify systems for which phonon screening is expected to be significant.

In summary, we generalized the {\it ab initio} Bethe-Salpeter equation approach to include both electronic and phonon contributions to the screened Coulomb interaction, $W$, and studied phonon screening effects on the electron-hole interactions in halide perovskites and other important semiconductors. We showed that {\it ab initio} BSE calculations including phonon screening can reduce the exciton binding energy of lead-halide perovskites significantly as compared to electronic screening alone,  reconciling two previous contradictory hypotheses on the importance of phonon screening in metal-halide perovskites.  We rationalized our results by generalizing the Wannier-Mott model for excitons in a phonon-screened environment. Within this model, we showed that phonon screening is important for other semiconductors, and can be traced back to four material specific parameters, $\mu$, $\omega_{\rm LO}$, $\varepsilon_{\infty}$ and $\varepsilon_0$. We derived a simple expression providing intuition for the importance of lattice vibrations on the excitonic properties of materials and outlined a general, simple, and quantitative approach to estimate the exciton binding energy correction using physical quantities that can be readily calculated theoretically or measured experimentally. By introducing a general framework to quantitatively account for phonon screening in {\it ab initio} BSE calculations, our study clarifies the importance of phonon screening corrections, and provides a necessary foundation for future treatment of polarons and higher order processes beyond two particle excitations for these and other complex materials.

\begin{acknowledgments}
The authors acknowledge A. Alvertis (Berkeley Lab), D. Qiu (Yale U), F. da Jornada (Stanford U), Z. Li (UC Berkeley), H. Paudial and R. Margine (SUNY Binghamton) for useful discussions. This work was supported by the Center for Computational Study of Excited-State Phenomena in Energy Materials (C2SEPEM) at Lawrence Berkeley National Laboratory, which is funded by the U. S. Department of Energy, Office of Science, Basic Energy Sciences, Materials Sciences and Engineering Division, under Contract No. DE-C02-05CH11231. MRF acknowledges support from the Engineering and Physical Sciences Research Council (EPSRC), grant no.  EP/V010840/1. The authors acknowledge computational resources provided by the National Energy Research Scientific Computing Center (NERSC) and Molecular Foundry, also supported by the Office of Science, Office of Basic Energy Sciences, of the US DOE under Contract DE-AC02-5CH11231. Additional computational resources were provided by the Extreme Science and Engineering Discovery Environment (XSEDE) supercomputer Stampede2 at the Texas Advanced Computing Center (TACC) through the allocation TG-DMR190070.
\end{acknowledgments}
%

\onecolumngrid
\newpage 

\begin{center}
	\large {\bf Phonon Screening of Excitons in Semiconductors: Halide Perovskites and Beyond \\
	Supplementary Information}
\end{center}

\renewcommand{\figurename}{Figure}
\renewcommand{\tablename}{Table}
\renewcommand{\thetable}{S\arabic{table}}
\renewcommand{\thefigure}{S\arabic{figure}}
\setcounter{figure}{0}
\setcounter{table}{0}

\section{I.~Computational Setup}
{\bf A. Ground state calculations}

For all calculations on lead halide perovskites, we use  experimental lattice parameters reported in Refs.~\cite{Linaburg2017,Sutton2018}, and relax the atomic positions. We study the low temperature orthorhombic phase, with lattice parameters summarized in Table~\ref{tb2}. All DFT calculations are performed using the generalized gradient approximation within the Perdew Burke Erzerhof parametrization (PBE)~\cite{Perdew}, including spin orbit coupling as implemented in the \textsc{Quantum Espresso} package~\cite{QE}. For all calculations we use the norm conserving fully relativistic pseudopotentials from the PseudoDojo database~\cite{Hamann2013,vanSetten2018},  with the following valence electrons configuration: Pb ($5d^{10}~6s^2~6p^4$), I ($5s^2~5p^5$), Br ($4s^2~4p^5$), Cl ($3s^23p^5$) and Cs ($5s^2~5p^6~6s^1$). We use a plane wave cutoff of 50~Ry and discretize the Brillouin zone using a half shifted Monkhorst-Pack grid of $6\times4\times6$, following the aspect ratio of the unit cell. 

For calculations on AlN, MgO, GaN and CdS we use PBE norm-conserving pseudopotentials from the Pseudo dojo database~\cite{vanSetten2018}. We use the lattice parameters from the Materials Project database. For these semiconductors we use a plane wave cutoff of 100 Ry and a shifted grid of $6\times6\times6$ to converge the electronic charge density. 

\vspace{0.2cm}
{\bf B. Quasiparticle eigenvalues}

We calculate the quasiparticle eigenvalues of the lead-halide perovskites with a one-shot $G_0W_0$ approximation as implemented in the BerkeleyGW code~\cite{BGW}. We calculate the static screened Coulomb interaction within the random-phase approximation (RPA)~\cite{Adler,Wiser}, and extend this dielectric function to finite frequencies using the Godby-Needs plasmon pole model~\cite{Godby}. We use a half-shifted $4\times4\times4$ $\mathbf{k}$-point mesh, a 14~Ry plane wave cutoff and 1000 bands to calculate the dielectric screening. In addition, we sum over 1000 bands in order to calculate the electronic self-energy on a $\Gamma$-centered $4\times4\times4$ $\mathbf{k}$-point mesh, following convergence studies reported in Ref.~\cite{Filip2014-2}, and used the static remainder approximation to optimize convergence with respect to the number of empty states~\cite{Deslippe2013}. These computational parameters are similar to previous $GW$ calculations on the quasiparticle band gap in halide perovskites, and are expected to yield band gaps converged within 0.1 eV~\cite{Filip2014-2, Filip2015, Davies2018, Sutton2018}.

$G_0W_0$ quasiparticle band gaps are underestimated with respect to experiment by up to 0.5~eV. This observation is consistent with several recent studies of the quasiparticle band structure of lead-halide perovskites, and can be attributed to the sensitivity of quasiparticle band gaps to the Kohn-Sham starting point~\cite{Leppert2019, Brivio2014, Filip2014-2, Scherpeltz2016, Wiktor2017}. One possible route to mitigate this underestimation is to perform quasiparticle calculations self-consistently, as discussed in Refs~\cite{Brivio2014, Filip2014-2}, or use a hybrid functional starting point, as reported in~\cite{Leppert2019}. 
Additionally, we note that in our calculations we  do not include semicore $d$ states for I and Br. As reported in Refs.~\cite{Filip2014-2,Scherpeltz2016}, quasiparticle band gaps calculated with and without semicore $d$ states for I or Br differ by up to 0.3~eV. Since this work is not focused on the study of quasiparticle band gaps, we do not include semicore $d$ states for Br and I, and calculate the band gap within the $G_0W_0$@PBE approximation in order to reduce the computational effort. 

We have separately converged the quasiparticle band gaps of GaN, AlN, CdS and MgO calculated within the $G_0W_0$ approximation, obtaining coverged quasiparticle band gaps within 0.1~eV with the following parameters summarized here for transparency: 
GaN ($4\times4\times4$ $\mathbf{q}$-point grid, 400 bands and 40 Ry polarizability cutoff), 
AlN ($6\times6\times6$ $\mathbf{q}$-point grid, 400 bands and 32 Ry polarizability cutoff),
CdS ($6\times6\times6$ $\mathbf{q}$-point grid, 500 bands and 40 Ry polarizability cutoff),
MgO ($6\times6\times6$ $\mathbf{q}$-point grid, 600 bands and 50 Ry polarizability cutoff). 

\vspace{0.2cm}
{\bf C. Exciton binding energies}

We calculate the exciton binding energy by solving the Bethe-Salpeter equation within the Tamm-Dancoff approximation, as implemented in the BerkeleyGW code~\cite{BGW}. We construct the electron-hole kernel on a $4\times4\times4$ $\mathbf{k}$-point grid, using 20 valence and 20 conduction bands, which is then interpolated to a fine mesh with 4 valence and 2 conduction bands (spin degenerate). It was previously shown in the case of MAPbI$_3$ that the exciton binding energy is sensitive to the density of the $\mathbf{k}$-point mesh used to diagonalize the BSE Hamiltonian, which constitutes the principal bottleneck for the calculation of exciton binding energies in halide perovskites~\cite{Bokdam2016}. In order to reach very dense $\mathbf{k}$-point meshes, we employ the patched sampling scheme, originally introduced in Ref.~\cite{Rohlfing1998}, whereby for the diagonalization of the BSE Hamiltonian we take into account only $\mathbf{k}$-points in a small patch around the $\Gamma$-point. The electron-hole interaction kernel and quasiparticle eigenvalues are interpolated on the fine grid using the method described in Ref.~\cite{Rohlfing2000,BGW}.

We first test convergence of the exciton binding energy with respect to the density of the $\mathbf{k}$-point mesh for the experimental cubic structure of CsPbBr$_3$, as reported in Ref~\cite{Moller1958} (Figure~S1). This structure is significantly simpler than the orthorhombic structures, with only 5 atoms in the unit cell. We use the patched sampling scheme in all these convergence tests with patches of 0.1~\AA$^{-1}$ (ie. we only consider $\mathbf{k}$-points which lie within 0.1~\AA$^{-1}$ of the band edge). We find that the exciton binding energy does not exhibit a linear dependence on the inverse number of $\mathbf{k}$-points, for equivalent full Brillouin zone grids ranging from $20\times20\times20$ up to $60\times60\times60$, in contrast with what was reported in prior work~\cite{Bokdam2016}. For this reason we chose not to make use of the extrapolation method proposed in Ref.~\cite{Bokdam2016}, and instead calculate directly the exciton binding energy using a dense grid of $\mathbf{k}$-points. We find that for the cubic CsPbBr$_3$ structure, the exciton binding energy is converged within 3~meV for a grid density of $60\times60\times60$. For all these convergence tests we do not calculate quasiparticle eigenvalues, and use the DFT eigenvalues instead.

We have also carefully checked the  convergence of the exciton binding with respect to the density of the  $\mathbf{k}$-point mesh and the size of the patch for the orthorhombic structures of the CsPbX$_3$ series. For all three compounds we use a mesh of $30\times30\times30$  centered at the $\Gamma$ point. We sample $\mathbf{k}$-points in this mesh which are within 0.15~\AA$^{-1}$ for CsPbI$_3$, 0.20~\AA$^{-1}$ for CsPbBr$_3$ and 0.35~\AA$^{-1}$ for CsPbCl$_3$. This grid density is consistent with the grid density calculated for the cubic structure, once the unit cell aspect ratio is taken into account. Based on our convergence tests, we find this computational setup is sufficient to converge exciton binding energies within 3~meV. Two valence and two conduction bands are sufficient to calculate exciton binding energies within the same level of convergence for all three perovskites. 

We use the same approach to calculate the exciton binding energies of GaN, and CdS, while for MgO and AlN we use the full $\mathbf{k}$-point mesh. In all exciton binding energy calculations we use 3 valence bands and one conduction band. We have separately converged the exciton binding energies with the density of the {$\mathbf{k}$}-point grid and the size, obtaining the following parameters summarized here for transparency:
GaN ($50\times50\times50$ $\mathbf{k}$-point grid, with a patch radius of 0.2~\AA$^{-1}$), 
AlN ($24\times24\times24$ $\mathbf{k}$-point grid, without patched sampling),
CdS ($60\times60\times60$ $\mathbf{k}$-point grid, with a patch radius of 0.1~\AA$^{-1}$),
MgO ($24\times24\times24$ $\mathbf{k}$-point grid, without patched sampling). 

\vspace{0.2cm}
{\bf D. Optical absorption spectrum}

We calculate the imaginary part of the dielectric function with ($\varepsilon_2^{\rm eh}$) and without ($\varepsilon_2^{\rm no~eh}$) taking into account electron-hole interactions, following~\cite{BGW, Rohlfing1998} as:
{\small
\begin{eqnarray}
	\varepsilon_2^{\rm eh}(\omega)  &=&   \frac{16\pi^2e^2}{\omega^2} 
	\sum_S | \mathbf{e}\cdot\langle 0 | \mathbf{v} | S \rangle |^2 \delta (\omega-\Omega_S) 
	\\ \nonumber	\varepsilon_2^{\rm no~eh}(\omega) &=&  \frac{16\pi^2e^2}{\omega^2}   
	\sum_{vc\mathbf{k}} | \mathbf{e}\cdot\langle v\mathbf{k} | \mathbf{v} | 
	c\mathbf{k}\rangle |^2 \delta(\omega-E_{c\mathbf{k}}+E_{v\mathbf{k}}). 
	\label{eq7}
\end{eqnarray}}
In order to reduce the computational effort, we approximate the velocity matrix elements as $\langle v\mathbf{k} | \mathbf{v} | c\mathbf{k}\rangle = -i \langle v\mathbf{k} | \nabla | c\mathbf{k}\rangle$, ignoring the non-local part of the Hamiltonian, as discussed in Ref.~\cite{BGW}. We have checked that this approximation only changes the magnitude of the absorption spectrum by up to 10\%. In Figure~1  of the main manuscript we plot the optical absorption coefficient as a function of energy, defined as $\alpha(\omega) = \omega \varepsilon_2(\omega)/cn(\omega)$, where $n(\omega)$ is the refractive index, approximated as a constant, as in Ref.\cite{Davies2018}. 

In order to capture all the features of the optical absorption spectrum we calculate the imaginary part of the dielectric function, using the eigenvalues of the BSE Hamiltonian, calculated for the entire Brillouin zone, rather than a patch, and using 4 valence and conduction bands. For this calculation we reduce the density of the $\mathbf{k}$-point grid to  $22\times22\times22$. The spectra shown in Figure~1 of the main manuscript were calculated using a Gaussian smearing with a constant width of 50 meV for all energies below the quasiparticle band gap, and an adaptive width which increases linearly from 50~meV to 82 meV below the quasiparticle band gap and remains constant at 82 meV above the band gap. We have checked that this choice of broadening does not impact the main features of the optical absorption spectrum; we use this approach for visualization purposes, so as to best reproduce the experimental spectra. 

\vspace{0.2cm}
{\bf E. Density functional perturbation theory calculations of dielectric constants}

We calculate the phonon dispersion spectrum, the high and low frequency dielectric constants, and the characteristic LO phonon frequency within density functional perturbation theory (DFPT), as implemented in the \textsc{Quanum Espresso} package. We use a half shifted $6\times4\times6$ {$\mathbf{k}$}-point mesh to calculate the ground state charge density, and interpolate through phonon frequencies calculated on a  {$\mathbf{q}$}-point grid of $2\times2\times2$ to obtain the phonon dispersion spectra shown in Figure~\ref{fig3}. For all three phonon calculations we do not include spin-orbit coupling effects. This choice is justified by previous computational studies of the vibrational properties of halide perovskites, which have explicitly shown that  inclusion of spin-orbit coupling does not impact the calculated vibrational spectrum~\cite{Perez2015}. In addition, we have checked that the characteristic LO phonon frequency reported in Table~1 of the main manuscript does not depend strongly on the polarization direction, and changes by less than 1~meV across different directions. Furthermore, the dielectric constants reported in Table~1 of the main manuscript are an isotropic average of the three diagonal elements of the dielectric constant tensor. 

\newpage

\section{II.~Methods} 
{\bf A. Quasiparticle eigenvalues}

We calculate the quasiparticle eigenvalues $E_{n\mathbf{k}}$ within a one-shot $G_0W_0$ approximation as~\cite{Hybertsen} 
{\small
\begin{equation}
	E_{n\mathbf{k}} = \epsilon_{n\mathbf{k}}+Z(\epsilon_{n\mathbf{k}}) \langle n\mathbf{k} 
	| \Sigma(\epsilon_{n\mathbf{k}}) - V_{\rm xc} | n\mathbf{k} \rangle,
	\label{eq1}
\end{equation}}
where $\epsilon_{n\mathbf{k}}$ are the DFT eigenvalues, $\Sigma(\omega)$ is the electron self energy operator, $V_{\rm xc}$ is the  exchange-correlation potential and $Z(\omega)$ is the quasiparticle renormalization factor expressed as $Z(\omega) = \Big[1-{\rm Re}(\partial \Sigma/ \partial \omega) \Big]^{-1}$~\cite{Hybertsen}. The electron self-energy is calculated in the $GW$ approximation as the convolution of the single particle Green's function, $G$ with the screened Coulomb interaction, $W$, defined below. The single particle Green's function is calculated starting from DFT as~\cite{Hedin}:
{\small
\begin{equation}\label{eq2}
	\displaystyle{G(\mathbf{r},\mathbf{r'};\omega) = \sum_{n\mathbf{k}}\frac{\psi_{n\mathbf{k}}(\mathbf{r})\psi^*_{n\mathbf{k}}(\mathbf{r'})}{\omega-\epsilon_{n\mathbf{k}}-i\eta}},
\end{equation}}
where the summation runs over occupied and unoccupied states, $\psi_{n\mathbf{k}}(\mathbf{r})$ is the DFT wave function corresponding to the energy eigenvalue $\epsilon_{n\mathbf{k}}$ and $\eta$ is an infinitesimally small constant, positive for occupied states and negative for unoccupied states~\cite{Hedin}. The screened Coulomb interaction is given by the expression  $W(\mathbf{r}, \mathbf{r'};\omega) = \varepsilon^{-1}(\mathbf{r}, \mathbf{r'}; \omega) v(\mathbf{r},\mathbf{r'})$, where $v(\mathbf{r},\mathbf{r'}) = 1/|\mathbf{r}-\mathbf{r'}|$ is the bare Coulomb potential and $\varepsilon(\mathbf{r}, \mathbf{r'};\omega)$ is the dielectric function. We calculate the frequency dependent dielectric function using the Godby-Needs plasmon pole model~\cite{Godby, Hybertsen}. 

\vspace{0.2cm}
{\bf B. Exciton binding energy}

We calculate the excitonic spectra by solving the Bethe-Salpeter equation (BSE) in the Tamm-Dancoff approximation
\cite{Rohlfing1998, Rohlfing2000}:
{\small\begin{equation}
      (E_{c\mathbf{k}}-E_{v\mathbf{k}}) A^S_{cv\mathbf{k}}+\sum_{c'v'\mathbf{k'}}
	K_{cv\mathbf{k};c'v'\mathbf{k'}}(\Omega_S) A_{c'v'\mathbf{k'}}^S= 
      \Omega_S A_{cv\mathbf{k}}^S,
      \label{eq4}
\end{equation}}
where $E_{c\mathbf{k}}$ and $E_{v\mathbf{k}}$ are the $GW$ quasiparticle energies corresponding to the unoccupied ($c$) and occupied ($v$) states, respectively at wave vector $\mathbf{k}$, and $K$ is the electron-hole interaction kernel, which consists of a frequency independent bare exchange term, $K^{\rm x}$, and a frequency dependent screened direct term, $K^{\rm D}$, written in Eq. 2 of the main manuscript in the single particle basis, and reproduced here for clarity~\cite{Strinati}: 
{\small\begin{align}\label{eq5}
        K^{\rm D}_{cv\mathbf{k},c'v'\mathbf{k'}}(\Omega) = -\Bigg \langle cv \mathbf{k}\Bigg | 
	\frac{i}{2\pi}\int d\omega~e^{-i\omega 0^+} W(\mathbf{r}, \mathbf{r'};\omega)
	\Bigg[\frac{1}{\Omega-\omega-(E_{c'\mathbf{k'}}-E_{v\mathbf{k}})+i0^+} + 
	\frac{1}{\Omega+\omega-(E_{c\mathbf{k}}-E_{v'\mathbf{k'}})+i0^+}\Bigg] & \Bigg| c'v' \mathbf{k'}\Bigg \rangle. 
\end{align}}
In Eq.~\ref{eq5}, $W(\mathbf{r}, \mathbf{r'};\omega)$ is the time ordered screened Coulomb interaction. Solving the BSE yields exciton energies, $\Omega_S$, and exciton expansion coefficients, $A_{cv\mathbf{k}}^S$.  The exciton binding energy is the difference between the lowest excitonic state and the quasiparticle band gap while the exciton wave function can be obtained as an expansion of free electron-hole pairs, within the Tamm-Dancoff approximation~\cite{Rohlfing1998, Rohlfing2000} as:
{\small
\begin{equation}
      |S\rangle = \sum_{cv\mathbf{k}} A^S_{cv\mathbf{k}} | cv \mathbf{k}\rangle. 
      \label{eq6}
\end{equation}}

As discussed in the main manuscript, the screened Coulomb  interaction $W$ can rigorously be expressed  as a sum of two contributions, an electronic, $W^{\rm el}$, and a phonon, $W^{\rm ph}$, contribution~\cite{Hedin1970}. Consequently, the electron-hole kernel naturally decouples into two terms, an electronic $K^{\rm el}$ and a phonon  $K^{\rm ph}$ term. Below, we derive an expression for the electron-hole kernel, including phonon screening.

\vspace{0.1cm}
{\it \underline{1. Electronic Contribution}}

	For context, we first sketch the derivation of the electronic contribution electron-hole kernel matrix elements, as detailed by Rohlfing and Louie~\cite{Rohlfing1998, Rohlfing2000}. Adopting a plasmon-pole model, we express the electronic contribution to the screened Coulomb interaction in real-space as~\cite{Rohlfing1998}:	
	{\small\begin{equation}\label{eq8}
		W^{\rm el}(\mathbf{r},\mathbf{r'};\omega) = \sum_{l}\frac{W_l(\mathbf{r}, \mathbf{r'})}{2}
		 \Bigg(\frac{\omega_l}{\omega-\omega_l+i0^+}- \frac{\omega_l}{\omega+\omega_l-i0^+}\Bigg).
	\end{equation}}
	Using this form for $W^{\rm el}(\mathbf{r}, \mathbf{r'};\omega)$ in Eq.~\ref{eq5}, the frequency integration can be performed analytically, and the direct kernel matrix elements becomes~\cite{Rohlfing1998}
	\begin{equation}\label{eq9}
		K^{\rm D}_{cv\mathbf{k};c'v'\mathbf{k'}} (\Omega_S) = -\sum_l \Bigg[ \frac{\omega_l}{2} 
		\Bigg \langle cv\mathbf{k} \Bigg | W_l(\mathbf{r},\mathbf{r'}) \Bigg| c'v'\mathbf{k'}\Bigg\rangle 
		\Bigg(\frac{1}{\Omega_S-E_{c'\mathbf{k'}}+E_{v\mathbf{k}}-\omega_l}+\frac{1}{\Omega_S-E_{c\mathbf{k}}+E_{v'\mathbf{k'}}-\omega_l}\Bigg)\Bigg].
	\end{equation}
	In our study we are interested in weakly bound excitons, for which $|\Omega_S-E_c+E_v| \ll \omega_l$. In this limit, we can rewrite Eq.~\ref{eq9} following ~\cite{Rohlfing1998} as:
	\begin{equation}\label{eq10}
		K^{\rm D}_{cv\mathbf{k};c'v'\mathbf{k'}} (\Omega_S) = -\Big \langle cv\mathbf{k} \Big | W^{\rm el}(\mathbf{r},\mathbf{r'},\omega = 0)
                \Big| c'v'\mathbf{k'}\Big\rangle.
	\end{equation}

{\it \underline{2. Phonon Contribution}}

    	The phonon contribution to the screened Coulomb interaction is mediated by the exchange of a virtual phonon, and can be written as~\cite{Hedin1970,Giustino2017}:
   	\begin{equation} \label{eq13}
        	W_{\rm ph}(\mathbf{r},\mathbf{r'}; \omega) =  \sum_{\mathbf{q} \nu} D_{\mathbf{q} \nu}(\omega) g_{\mathbf{q}\nu}(\mathbf{r})g_{\mathbf{q} \nu}^*(\mathbf{r'}),
    	\end{equation}
    	where $D_{\mathbf{q}\nu }(\omega)$ is the time ordered phonon propagator and defined as 	$\displaystyle{D_{\mathbf{q} \nu}(\omega) = \frac{1}{\omega - \omega_{\mathbf{q} \nu}+i0^+}} -  \frac{1}{\omega + \omega_{\mathbf{q} \nu}-i0^+}$ and $g_{\mathbf{q}\mathbf{\nu}}(\mathbf{r})$ is the electron-phonon vertex associated with the wave-vector $\mathbf{q}$ and mode $\nu$, written in real space. 
        
        Replacing Eq.~\ref{eq13} into Eq.~\ref{eq5}, we find the following expression for the phonon contribution to the kernel matrix elements, written in the single particle basis:
        \begin{equation}\label{kphcv}
            K^{\rm ph}_{cv\mathbf{k},c'v'\mathbf{k'}}(\Omega) = -\sum_{\mathbf{q}\nu} g_{cc'\nu}(\mathbf{k'},\mathbf{q})g^*_{vv'\nu}(\mathbf{k'},\mathbf{q}) 
            \Bigg[\frac{1}{\Omega-(E_{c\mathbf{k}}-E_{v'\mathbf{k'}})-\omega_{\mathbf{q}\nu}+i\eta}+\frac{1}{\Omega-(E_{c'\mathbf{k'}}-E_{v\mathbf{k}})-\omega_{\mathbf{q}\nu}+i\eta}\Bigg]. 
        \end{equation}
        Eq.~\ref{kphcv} can be written equivalently in the exciton basis, using Eq.~\ref{eq6}, which we will use in the following to extract more physical insight. 
        \begin{eqnarray}\label{kexc}
            K^{\rm ph}_{SS'}(\Omega) = -\sum_{cv\mathbf{k}c'v'\mathbf{k'} \nu} 
            A^{S*}_{cv\mathbf{k}}g_{cc'\nu}(\mathbf{k'},\mathbf{q})g_{vv'\nu}^*(\mathbf{k'},\mathbf{q})A_{c'v'\mathbf{k'}}^{S'} \times \\ \nonumber  \bigg[\frac{1}{\Omega-(E_{c\mathbf{k}}-E_{v'\mathbf{k'}})-\omega_{\mathbf{q}\nu}+i\eta}+\frac{1}{\Omega-(E_{c'\mathbf{k'}}-E_{v\mathbf{k}})-\omega_{\mathbf{q}\nu}+i\eta}\bigg]. 
        \end{eqnarray}
    In this work, we make the approximation that the matrix  $K^{\rm D; ph}(\Omega_S)$ is diagonal, and calculate the correction as $\Delta \Omega_S = {\rm Re} \langle S | K^{\rm D; ph}(\Omega_S) | S \rangle$.

        As mentioned in the main manuscript, we proceed with two additional approximations. 
        Firstly, we calculate the electron-phonon vertex appearing in Eqs.~\ref{eq13} and \ref{kexc} using the {\it ab initio} generalization of the Fr\"ohlich model (written in reciprocal space, in atomic units) introduced in Ref.~\cite{Verdi2015}:
        \begin{equation}\label{verdi}
        g_{\mathbf{q}\nu}=i\frac{4\pi}{V}\sum_{\kappa}\Bigg(\frac{1}{2NM_\kappa\omega_{\mathbf{q}\nu}}\Bigg)^{1/2}\frac{\mathbf{q}\cdot\mathbf{Z_{\kappa}}\cdot\mathbf{e}_{\kappa\nu}(\mathbf{q})}{\mathbf{q}\cdot\mathbf{\varepsilon_\infty}\cdot\mathbf{q}},
        \end{equation}
        where $V$ is the unit cell volume, $\kappa$ indexes the atom, $N$ is the number of unit cells, $M_{\kappa}$ are the atomic masses, $Z_{\kappa}$ are the Born effective charges, and $\mathbf{e}_{\kappa\nu}(\mathbf{q})$ are the phonon eigenvectors associated with the phonon frequencies $\omega_{\nu\mathbf{q}}$. This approximation is justified given that phonon screening effects are expected to arise predominantly due to long range contributions of optically active phonon modes. Secondly, we parametrize the exciton wave function $A_{cv\mathbf{k}}$ with the hydrogenic model expression (for the first excited state only), $\displaystyle{A_{\mathbf{k}} = \frac{(2a_0)^{3/2}/\pi}{(1+a_0^2k^2)^2}}$. This approximation is justified in Figure~1 of the main manuscript, where we show explicitly that the hydrogen model yields an accurate description of the exciton wave function in lead-halide perovskites. With these two approximations, we can rewrite Eq.~\ref{kexc} as:
        \begin{equation}
            \Delta\Omega_S = -\frac{8 a_0^3}{\pi^2} \sum_{\mathbf{kq}\nu} \frac{|g_{\mathbf{q}\nu}|^2 }{[1+a_0^2|\mathbf{k}|^2]^2[1+a_0^2|\mathbf{k}+\mathbf{q}|^2]^2}  
 \bigg[\frac{1}{\Omega_S - \Delta_{c\mathbf{k}v'\mathbf{k+q}}-\omega_{\mathbf{q}\nu}} + \frac{1}{\Omega_S-\Delta_{c'\mathbf{k+q}v\mathbf{k}}-\omega_{\mathbf{q}\nu}} \bigg],
        \end{equation}
        We use this expression to calculate the phonon screening correction to the exciton binding energy, using as input the exciton energy as obtained from a standard BSE calculation, the quasiparticle energies from $G_0W_0$, and the phonon frequencies, eigenvectors and Born effective charges from DFPT.  

        To be able to observe a general trend, we make some further simplifications to Eq.~\ref{kexc} through the following well justified assumptions:
        
        \noindent (i) A single dispersionless phonon contributes most significantly to the phonon kernel, such that the electron-phonon vertex can be replaced by the Fr\"ohlich model expression,~\cite{Frohlich}
	    \begin{equation}\label{eq14}
        	g^{\rm F}_{\mathbf{q}}(\mathbf{r}) = \frac{i}{|\mathbf{q}|} \Bigg[\frac{4 \pi}{N V} \frac{\omega_{LO}}{2} \Bigg(\frac{1}{\varepsilon_{\infty}} - \frac{1}{\varepsilon_0} \Bigg) \Bigg]^{\frac{1}{2}} e^{i \mathbf{q} \cdot \mathbf{r}}.
    	\end{equation}
    	\noindent (ii) The electronic bands are parabolic, with equal electron and hole effective masses, and in the limit of small $\mathbf{q}$ we have $\displaystyle{\Delta_{c\mathbf{k}v'\mathbf{k+q}} \sim E_g + \frac{|\mathbf{k}|^2}{2\mu} ~ \sim E_g + \frac{|\mathbf{k+q}|^2}{2\mu}}$.
        Within these approximations, and using the identity that $a_0 = 1/(2E_B\mu)^{1/2}$, Eq.~\ref{kexc} becomes:
                \begin{eqnarray}\label{eqsum}
            \Delta\Omega_S = \frac{8 a_0^3}{\pi^2}\frac{4\pi}{2 NV} \frac{\omega_{LO}}{E_B+\omega_{LO}}\Big(\frac{1}{\varepsilon_\infty}-\frac{1}{\varepsilon_0}\Big)
            \sum_{\mathbf{kq}} \frac{1}{[1+a_0^2|\mathbf{k}|^2]^2[1+a_0^2|\mathbf{k}+\mathbf{q}|^2]^2}  \frac{1}{|\mathbf{q}|^2} \times \\ \nonumber
 \bigg[\frac{1}{1 + |\mathbf{k}|^2/[2\mu(\omega_{LO}+E_B)]} + \frac{1}{1 + |\mathbf{k}+\mathbf{q}|^2/[2\mu(\omega_{LO}+E_B)]} \bigg].
        \end{eqnarray}
 For simplicity, we introduce the notation $b_0^2 = 1/[2\mu(\omega_{LO}+E_B)]$. By taking advantage of the symmetry of Eq.~\ref{eqsum}, we can further simplify this expression to:
 \begin{eqnarray}\label{eqsum2}
            \Delta \Omega_S = \frac{4\pi}{ NV}\frac{\omega_{LO}}{E_B+\omega_{LO}}\Big(\frac{1}{\varepsilon_\infty}-\frac{1}{\varepsilon_0}\Big)
            \sum_{\mathbf{kq}} \frac{(2a_0)^{3/2}/\pi}{[1+a_0^2|\mathbf{k}|^2]^2}\frac{(2a_0)^{3/2}/\pi}{[1+a_0^2|\mathbf{k}+\mathbf{q}|^2]^2}  \frac{1}{|\mathbf{q}|^2} \frac{1}{1 + b_0^2|\mathbf{k}|^2}.
        \end{eqnarray}
To proceed, we use the following two identities:
\begin{equation}\label{four1}
     \frac{1}{(2\pi)^{3/2}} \int d\mathbf{r} e^{i(\mathbf{k+q})\cdot\mathbf{r}}F(\mathbf{r}) = \frac{(2a_0)^{3/2}/\pi}{[1+a_0^2|\mathbf{k+q}|^2]^2} , 
\end{equation}
and 
\begin{equation}\label{four2}
        \frac{1}{(2\pi)^{3/2}} \int d\mathbf{r} e^{-i\mathbf{k}\cdot\mathbf{r}}G(\mathbf{r}) = \frac{(2a_0)^{3/2}/\pi}{[1+a_0^2|\mathbf{k}|^2]^2[1+b_0^2|\mathbf{k}|^2]}.
\end{equation}
Substituting Eq.~\ref{four1} and \ref{four2} into \ref{eqsum2}, and using the identities $\displaystyle{\frac{1}{r} = \frac{4\pi}{NV} \sum_\mathbf{q} \frac{1}{|\mathbf{q}|^2}e^{i\mathbf{q}\mathbf{r}}}$ and $\displaystyle{\delta(\mathbf{r}-\mathbf{r'}) = \frac{1}{(2\pi)^3} \sum_{\mathbf{k}}e^{i\mathbf{k}\cdot(\mathbf{r-r'})}}$, with $\delta(x)$ the Dirac delta function, we find:
\begin{equation}\label{eqint}
    \Delta\Omega_S = \frac{\omega_{LO}}{E_B+\omega_{LO}}\Big(\frac{1}{\varepsilon_\infty}-\frac{1}{\varepsilon_0}\Big) \int d\mathbf{r} G(\mathbf{r}) F(\mathbf{r}) \frac{1}{r}.            
\end{equation}
The function $F(\mathbf{r})$ is the hydrogenic wave function expressed in real space as $F(\mathbf{r}) = \exp(-r/a_0)/(\pi a_0^3)^{1/2}$, and $G(\mathbf{r})$ can be calculated as the inverse Fourier transform of the right hand side of Eq.~\ref{four2}, obtaining $\displaystyle{G(\mathbf{r}) = \frac{a_0^{1/2}}{\pi^{1/2} (a_0^2-b_0^2)^2r}\Big[-2a_0b_0^2(e^{-r/a_0}-e^{-r/b_0})+(a_0^2-b_0^2)r e^{-r/a_0}\Big]}$. With these two expressions, the integral in Eq.~\ref{eqint} yields a simple expression for the phonon screening correction to the exciton binding energy as a function of the bare exciton binding energy, LO phonon frequency, and the static and dynamic dielectric constants, $\varepsilon_{\infty}$ and $\varepsilon_0$:
\begin{equation}
   \Delta E_B = -\Big(\frac{1}{\varepsilon_\infty}-\frac{1}{\varepsilon_0}\Big) \frac{\omega_{LO}}{\omega_{LO}+E_B} \frac{(a_0+3b_0)a_0}{(a_0+b_0)^3}.
\end{equation}
Substituting $a_0$ and $b_0$ as given above, we obtain Eq.~7 of the main manuscript. Furthermore, if we neglect electron dispersion, as discussed in the main manuscript, then the parameter $b_0$ vanishes, leading to the simpler (but less accurate) expression given in the main manuscript. 

\bibliography{bibliography}

\newpage 

\begin{table*}[h!]
	\begin{center}
		\begin{tabular}{l r r r r r r r r r r}
			\hline
			\hline
			           & & $a$ (\AA) & & $b$ (\AA) & & $c$ (\AA) & & Space Group 
				   & & Ref.\\
			\hline
			CsPbCl$_3$ & & 7.902 & & 11.248 & & 7.899 & & $Pnma$ 
			           & & \cite{Linaburg2017}\\
			CsPbBr$_3$ & & 8.250 & & 11.753 & & 8.204 & & $Pnma$ 
			           & & ~\cite{Linaburg2017}\\
			CsPbI$_3$  & & 8.856 & & 8.576  & & 12.472 & & $Pbnm$ 
				   & & \cite{Sutton2018}\\
			\hline
			\hline
		\end{tabular}
		\caption{\small Experimental lattice parameters and space groups of the orthorhombic phases of the three halide perovskites studied in this work. Complete structural information can be found in the references cited in the table.\label{tb2}}
	\end{center}
\end{table*}

\begin{table*}[h!]
	\begin{center}
		\begin{tabular}{ l  cc cc cc  }			
\hline
\hline
			Compound	  && DFT && $G_0W_0$ && Exp. \\ 
			\hline
			CsPbI$_3$        && 0.85 && 1.46 && 1.7~\cite{Yang2017} \\
			CsPbBr$_3$        && 0.95 && 1.85 && 2.4~\cite{Yang2017}  \\
			CsPbCl$_3$         && 1.35 && 2.74 && 3.0~\cite{Heindrich1978}    \\
\hline
\hline
		\end{tabular}
		\caption{\small Summary of band gaps calculated within DFT+SOC and $G_0W_0$@PBE + SOC in this work, and compared with experimental band gaps reported in Refs.~\cite{Heindrich1978, Yang2017}.\label{tb3}}
	  	\end{center}
	  	
\end{table*}

\begin{table*}[h!]
	   \begin{center}
		\begin{tabular}{ l  c c c c c c c  }			
\hline
\hline
			& \multicolumn{3}{c}{Hole Effective Masses} & & 
			\multicolumn{3}{c}{Electron Effective Masses}\\
\hline			
			& CsPbI$_3$ & CsPbBr$_3$ &  CsPbCl$_3$ & 
 			& CsPbI$_3$ & CsPbBr$_3$ &  CsPbCl$_3$ \\
\hline
			$m_1$ &	0.25 (0.24) & 0.22 (0.23) & 0.28 (0.31) && 0.17 (0.16)  & 0.21 (0.21) & 0.29 (0.30) \\
			$m_2$ & 0.22 (0.18) & 0.18 (0.18) & 0.25 (0.25) &&  0.22 (0.19) & 0.18 (0.18) & 0.27 (0.28) \\
			$m_3$ & 0.19 (0.18) & 0.21 (0.21) & 0.26 (0.28) &&  0.22 (0.18) & 0.21 (0.21) & 0.29 (0.30)\\
			\\[-0.2cm]
			Average & 0.22 (0.20) & 0.20 (0.20) & 0.26 (0.28) && 0.20 (0.18) & 0.20 (0.21) & 0.29 (0.30) \\
\hline
\hline
		\end{tabular}
		\caption{\small Summary of $G_0W_0$ effective masses calculated the interpolation method described in Ref.~\cite{Deslippe2013} (and Wannier interpolation as described in Ref.~\cite{Filip2015}). In Figure~1 of the main manuscript we use the Wannier interpolated values to calculate the hydrogen model exciton binding energies.\label{tb3}}
	  	\end{center}
\end{table*}

\begin{table*}[h!]
\begin{center}
		\begin{tabular}{ l c c c c c c c c c c c c c c c }			
\hline
\hline
			& $E_{\rm g}^{\rm G_0W_0}$ & & $E_{\rm g}^{\rm exp}$ & & $\varepsilon_\infty^{GW}$ & & $\varepsilon_\infty^{\rm DFPT}$ & $\varepsilon_0^{\rm DFPT}$ 
			& $ \omega_{\rm LO}$ 
			& & $E_{\rm B}$ & &  $\Delta E_B$ & & 
			$E_{\rm B}^{\rm exp.}$ \\ 
		& (eV) & & (eV) & & & & & & (meV) & & (meV) & & (meV) & & (meV)\\
\hline			
			CsPbCl$_3$ & 2.74 & & 3.0~\cite{Heindrich1978} & & 3.7 & & 3.8 & 17.5 & 26 & & 146 & & -19 (-17) & & 
			72$\pm$3~\cite{Zhang2016}, 64$\pm1.5$~\cite{Baranowski2020-1}\\
			CsPbBr$_3$ & 1.85 & & 2.4~\cite{Yang2017} & & 4.5 & & 4.4 & 18.6 & 18 & & 70  & & -12 (-12) & &  
			33$\pm 1$~\cite{Yang2017},38$\pm 3$~\cite{Zhang2016} \\
			CsPbI$_3$  & 1.46 & & 1.7~\cite{Yang2017} & & 5.5 & & 5.4 & 22.5 & 14 & & 48  & & -9 (-8) & &
			15$\pm$1~\cite{Yang2017}\\
			\\[-0.2cm]
			GaN        & 2.84 & & 3.5~\cite{Madelung} & & 5.7 & & 5.8  & 10.7 & 87  & & 60 & &  
    				-22 (-27) & & 
			20~\cite{Muth1997, Madelung}\\ 
			AlN        & 5.60 & & 6.1~\cite{Madelung} & & 4.4 & & 4.4  & 8.4  & 112 & & 149 & & -37 (-38) & & 
			48~\cite{Leute2009}, 80~\cite{Li2003}\\
			MgO        & 7.10 & & 7.7~\cite{Madelung} & & 3.3 & & 3.2  & 10.6 & 85.6  & & 370 
				& & -52 (-50) & & 80~\cite{Walker1968}, 145$\pm$20~\cite{Whited1973}\\
			CdS      & 2.27 & & 2.5~\cite{Madelung} &  & 5.9 & & 6.2  & 10.6  & 34  & & 41 & & -8 (-9)  & & 
			    28~\cite{Jakobson1994}, 30.2~\cite{Voigt1979}\\
\hline
\hline
		\end{tabular}
		\caption{\small Summary of the static ($\varepsilon_0$) and high frequency dielectric ($\varepsilon_\infty$) calculated from RPA and DFPT, the energy of the highest frequency phonon mode, $ \omega_{\rm LO}$ calculated from DFPT, the high frequency, the bare exciton binding energy, and the phonon screening correction calculated using Eq.(7) (and via the {\it ab initio} Fr\"ohlich vertex), as discussed in the main manuscript.  
		\label{tb4}}
	  	\end{center}
\end{table*}

\newpage

\begin{figure*}[h!]
	\begin{center}
		\includegraphics[width=0.5\textwidth]{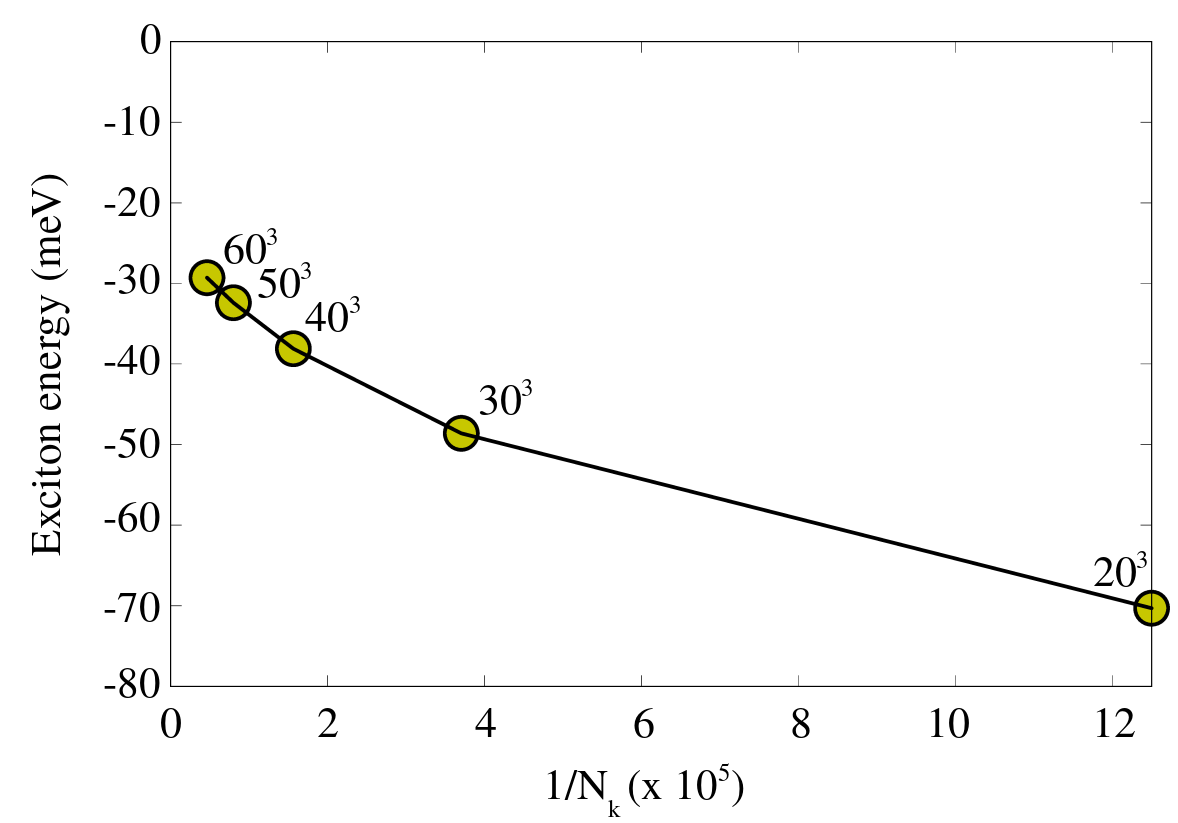}
		\caption{Convergence of the exciton binding energy with the density of the $\mathbf{k}$-point mesh. On the plot $N_k$ is the total number of $\mathbf{k}$-points in the mesh. Calculations are performed on the cubic CsPbBr$_3$ using a small patches around the valence band top and conduction band bottom,  with a radii of 0.1~bohr$^{-1}$. 
	\label{fig0}}
	\end{center}
\end{figure*}

\begin{figure*}[h!]
	\begin{center}
		\includegraphics[width=0.75\textwidth]{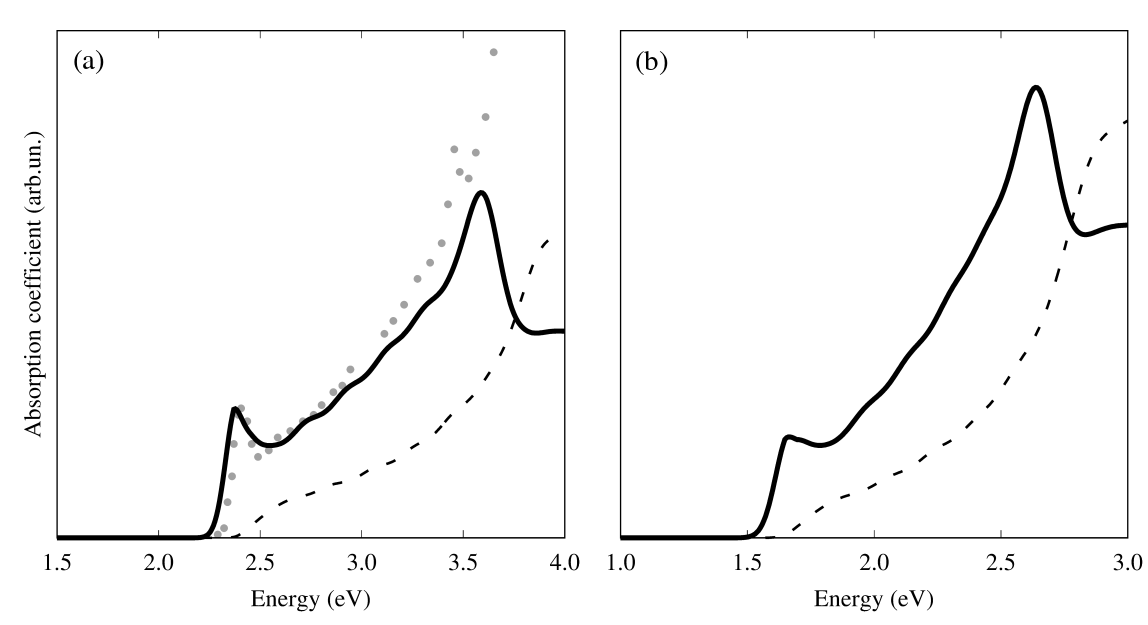}
		\caption{Comparison between optical absorption spectra calculated within GW/BSE (continuous black line) and the independent particle approximation (dashed black line) for CsPbBr$_3$ (a) and CsPbI$_3$ (b), respectively. Calculated optical spectra are blueshifted by 0.6 eV and 0.25 eV for CsPbBr$_3$ and CsPbI$_3$, respectively, in order to match experimental optical band gaps. Grey disks in (a) correspond to the experimental measurement reported in Ref.~\citenum{Heindrich1978}.
	\label{fig1}}
	\end{center}
\end{figure*}

\begin{figure*}[h!]
	\begin{center}
		\includegraphics[width=0.5\textwidth]{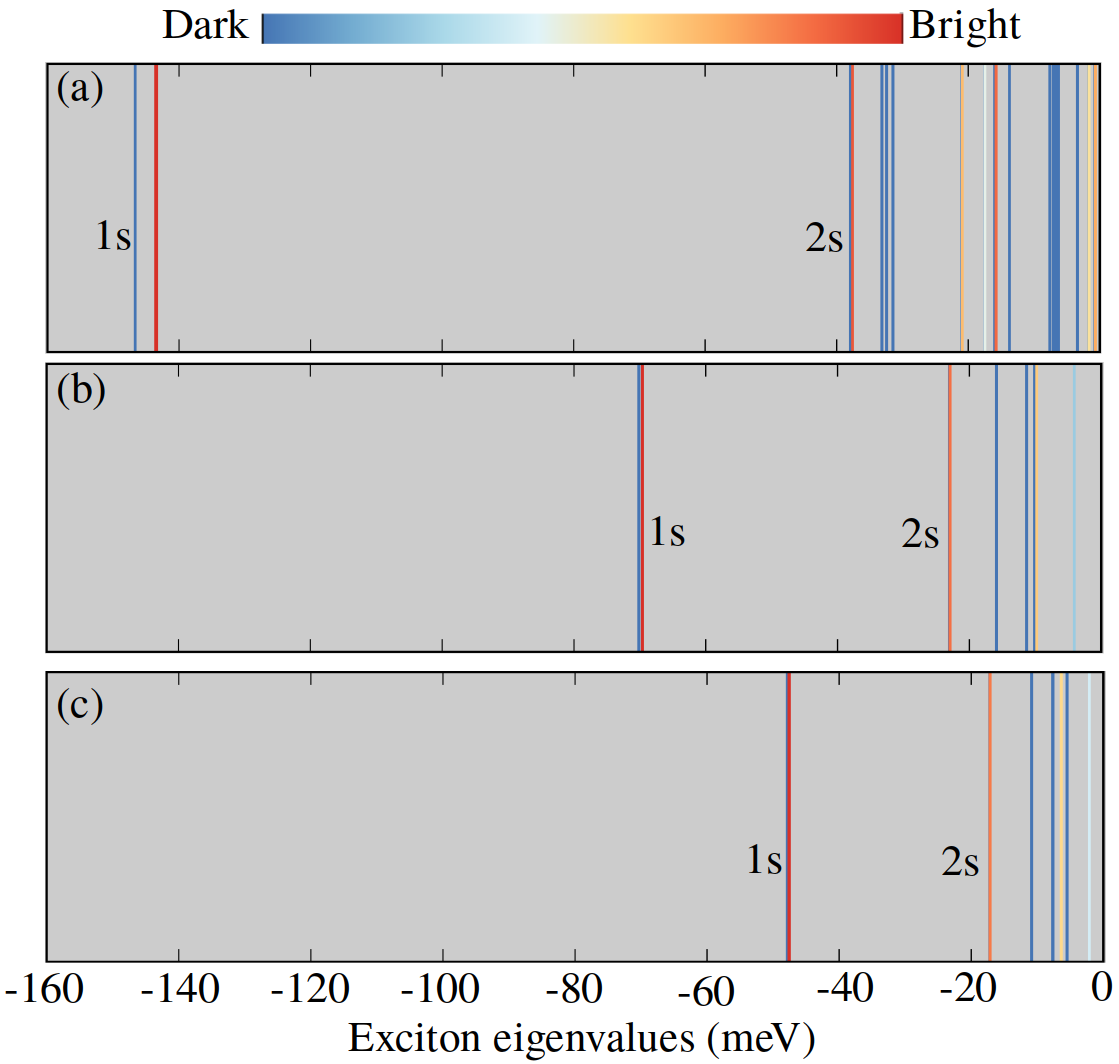}
		\caption{Bar plot of the exciton fine structure for CsPbX 3 , X = Cl (a), Br (b) and I (c). The color of the bar corresponds to the strength of the dipole matrix element. The 1s and 2s excitonic states are marked directly on the plot.\label{fig2}}
	\end{center}
\end{figure*}

\begin{figure*}[h!]
	\begin{center}
		\includegraphics[width=1.0\textwidth]{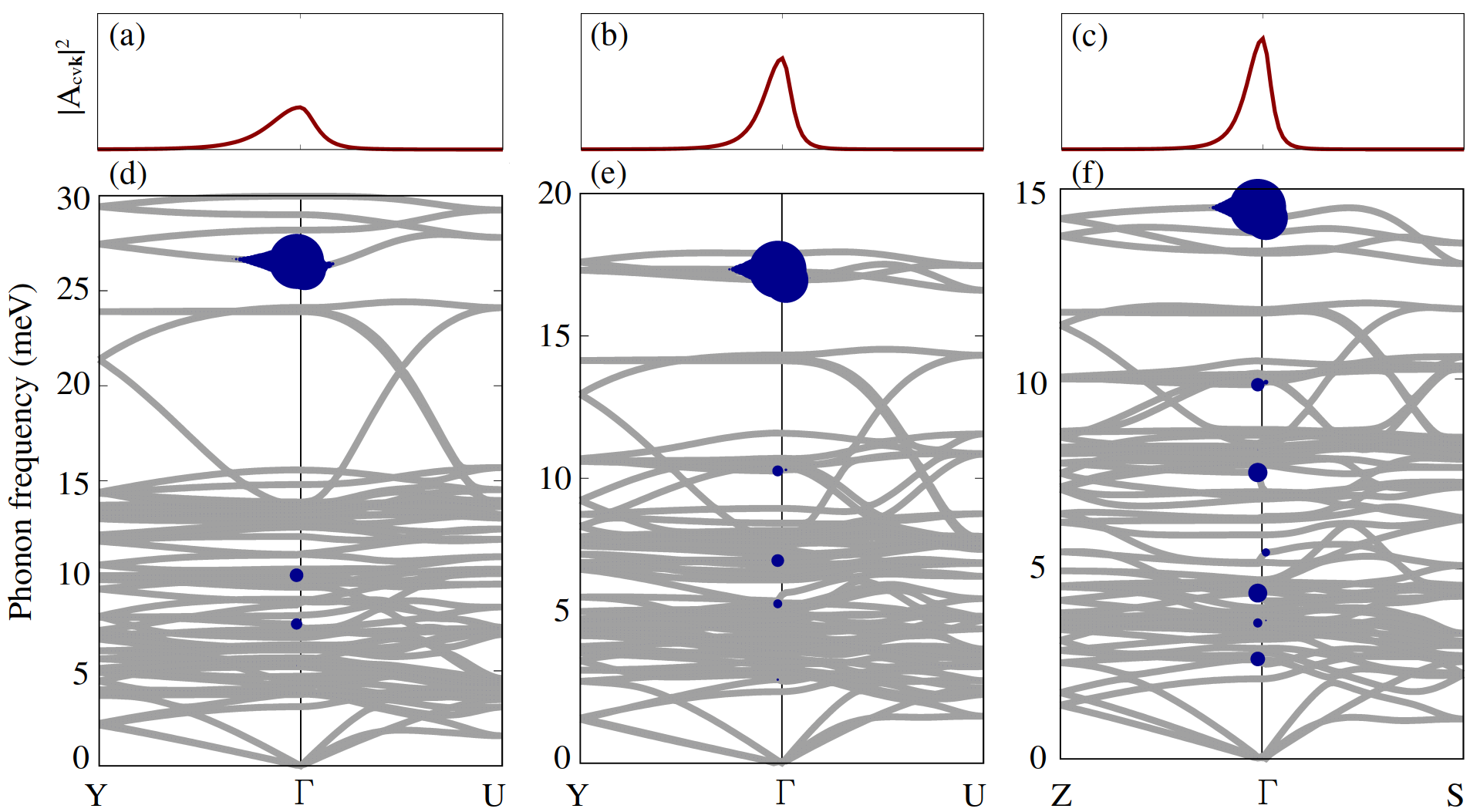}
		\caption{
		Exciton wave functions plotted along the high symmetry path for CsPbCl$_3$ (a), CsPbBr$_3$ (b) and CsPbI$_3$ (c). Phonon decomposition of the phonon screening correction computed using the {\it ab initio} Fr\"ohlich vertex (dark blue circles) overlayed over the phonon dispersion spectrum  CsPbCl$_3$ (d), CsPbBr$_3$ (e) and CsPbI$_3$ (f) from DFPT. The size of the points is proportional the phonon screening correction, in logarithmic scale
		 \label{fig3}}
	\end{center}
\end{figure*}

 \end{document}